# Static analysis of executable files by machine learning methods

Prudkovskiy Nikolay[1][2]

**Abstract**

The paper describes how to detect malicious executable files based on static analysis of their binary content. The stages of pre-processing and cleaning data extracted from different areas of executable files are analyzed. Methods of encoding categorical attributes of executable files are considered, as are ways to reduce the feature field dimension and select characteristic features in order to effectively represent samples of binary executable files for further training classifiers. An ensemble training approach was applied in order to aggregate forecasts from each classifier, and an ensemble of classifiers of various feature groups of executable file attributes was created in order to subsequently develop a system for detecting malicious files in an uninsulated environment.

## Introduction

In order to frustrate efforts to analyze malware and create signatures to identify viruses, the latest viruses are using polymorphism and metamorphism on an increasingly frequent basis. This means that the number of variations within malware families is constantly growing, which poses a serious problem for the developers of antivirus products. Approaches based on searching for signatures in files are no longer effective. They are being replaced by the dynamic analysis of malicious code in an isolated environment [1], as well as the use of various heuristic detection methods [2].

This paper proposes a method for detecting malicious portable executable (PE) files on Windows operating systems based on the static analysis of their binary contents. This operating system has been chosen due to it immense popularity, which inevitably results in a large number of malicious files being created specifically for this OS.

[1] Developer of machine learning algorithms, Group-IB. Email: prudkovskiy@group-ib.com
[2] Student of Bauman Moscow State Technical University, 6 course of study, Department IC8 «Information Security». Email: pns15u366@student.bmstu.ru

There are many approaches to the static analysis of executable files. The most popular of them are: byte n-gram analysis [3], classification of file types based on byte frequency analysis [4], analysis based on n-grams of opcodes [5,6], analysis based on state machines that detect anomalies in the code [7], detection of new malicious files based on string attributes [8, 9], classification based on attributes extracted from PE file structure [10].

We would like to note that the central stage of the malware detection process is the choice of PE file presentation, as well as the generation and selection of informative features. At this stage, it is necessary to obtain the fullest representation of the attribute space, which facilitates the efficient identification of malicious executables.

This paper poses the problem of developing a machine learning module that can classify PE files as malicious or legitimate with a high level of accuracy before they can run in a non-insulated environment. The most important objective at this stage is minimizing type I errors (false positives), caused by an imbalance of file classes (legitimate software being more prevalent). The engine design algorithm should not depend on the choice of platform on which it will subsequently work in field conditions.

The dataset being analyzed consists of 716,000 malicious files. It was obtained from "Warehouse", which is a centralized repository of malicious files collected by Group-IB using its proprietary TDS Polygon tool and third-party sandboxes. A set of 69,100 clean files was collected from various versions of Windows OS (Win7-32, Win7-64, and Win10-64), as well as by installing the following additional legitimate software:

- Instant messengers;
- Remote conference facilities;
- Office applications from different vendors;
- ERP and CRM systems;
- Text editors;
- Graphical editors;
- Web browsers;
- Mail clients;
- Security tools;
- Computer games;

- Antivirus protection tools;
- Virtualization tools;
- Servers (DB, mail, etc.)

# 1. Data preparation

Data preparation and cleaning are important tasks that must be completed before the dataset can be used to train the model. Raw data is often distorted and unreliable, and some values may be missing. The use of raw data in modeling may lead to inaccurate results.

## 1.1. The main objectives of data preparation

If any problems with the dataset are discovered, the data must be prepared and cleaned, which often involves removing missing values, normalizing data, discretizing, and processing text values to remove or replace embedded characters that can have a negative impact on the data type.

- *Data cleaning*: filling in missing values, detecting and removing noisy data and outliers.
- *Data transformation*: normalizing data to reduce volume and noise.
- *Data compaction*: selecting a subset of data or attributes to simplify data processing.
- *Data discretization*: converting continuous data attribute values into a finite set of intervals to facilitate the use of specific machine learning methods.

## 1.2. Data cleaning

When dealing with missing values, the first step that should be taken is to determine their cause; this will help in finding a solution to the problem and prevent data with missing values from being processed when it is known to be erroneous.

Methods for processing missing values [11]:

- *Deletion* is the simplest and most data-intensive method, whereby the entire record with a missing value is deleted. The main disadvantage of this approach is that if the record contains other data in addition to the missing value, then information required for training the model is also lost when the record is deleted.
- *Substitution of dummy data* involves replacing missing values with dummy ones: for example, substituting "unknown" for categorical values, or zeroes for numerical ones.
- *Mean imputation* – the missing numerical data can be replaced by the mean of the variable value.

- *Most frequently used value imputation* – missing categorical values can be replaced by the most frequently used value.
- *Regression imputation* – use of the regression method to replace missing values with regression values.

The last way of dealing with missing data is worth discussing in more detail. In regression analysis, the regression coefficients are essentially calculated based on the observed values that do not contain missing data, and the missing value of the dependent variable is calculated and restored based on these calculations. [12]

There are two problems that can be identified in this approach. The first is that due to the nature of regression, random fluctuations are excluded completely. This, for example, is what causes the same reconstructed values to be imputed into different observations for the same set of independent variable values. This means that in datasets with a large percentage of cases with missing values, there is a very noticeable bias towards mean values in the results. A random imputation method is used in order to compensate for this, where random values are added to the calculated value, e.g. taken from the regression equation residuals obtained on complete datasets. The second problem is that by using a set of independent variables in the regression equation which is too large, we risk modeling noise instead of certain meaningful variable values.

The maximum likelihood estimation (MLE) method arose from these two problems, which is specifically designed for dealing with missing information in large subsets.

The method is based on the following assumption: what "happened" in the study is what should have happened, i.e. the events that actually took place had the highest probability for the system under review, and were simultaneously the most relevant to the research tool being used. Therefore, all the unknown data that we are trying to recover should be sought in such a way that it is in the best possible agreement with the data that is already present in the database. This allows "the most plausible" estimates of the missing data to be made.

To use more formal language, we can say that the maximum likelihood estimate of an $\hat{\theta}$ unknown parameter $\theta$ is the point $\tilde{\theta}$ at which $f(X, \theta)$ a function is maximized (as a function of $\theta$ for fixed $X_1, \ldots, X_n$)

$$\hat{\theta} = arg \max_{\theta} f(X, \theta),$$

where $f(X, \theta) = f_\theta(X_1) \times f_\theta(X_2) \times ... \times f_\theta(X_n) = \prod_{i=1}^{n} f_\theta(X_i)$ is called a likelihood function. [13]

### 1.3. Data normalization

Normalizing data means scaling the numerical values into a specified range. The following are common data normalization methods. [14]

**Minmax normalization**

Linear data transformation within the range, e.g. from 0 to 1, where the minimum and maximum scalable values correspond to 0 and 1, respectively.

$$X_{norm} = \frac{X - X_{min}}{X_{max} - X_{min}}$$

**Z-score normalization**

Data scaling based on the mean and standard deviation: the difference between data point value and mean is divided by the standard deviation.

$$X_{norm} = \frac{X - \mu}{\sigma}$$

### 1.4. Dimensionality reduction

There are various methods for reducing the dimension of the data space without significantly compromising the information content of the dataset. Dimension reduction not only helps speed up model training, but in most cases it also helps filter out some noise and unnecessary details, thereby ensuring higher performance.

In addition to speeding up learning, dimensionality reduction is also extremely useful in data visualization. Reducing the number of dimensions to two or three allows high dimensionality training sets to be represented graphically, which often offer important insights, as they allow patterns such as clusters to be visually detected.

One of the most popular dimensionality reduction algorithms is Principal Component Analysis (PCA) [15]. First, it identifies the hyperplane that is closest to the data, and then projects the data onto it.

Before the training set can be projected onto a hyperplane with fewer dimensions, the correct hyperplane needs to be selected. Figure 1 on the left shows a simple two-dimensional dataset with three different axes (one-dimensional hyperplanes). On the right, the result of projecting the dataset onto each axis is shown. It can be seen that the projection onto the solid line preserves the maximum variance of the dataset, while the projection onto the dotted line removes the variance almost completely (the projection onto the dashed line shows the intermediate value of the preserved variance).

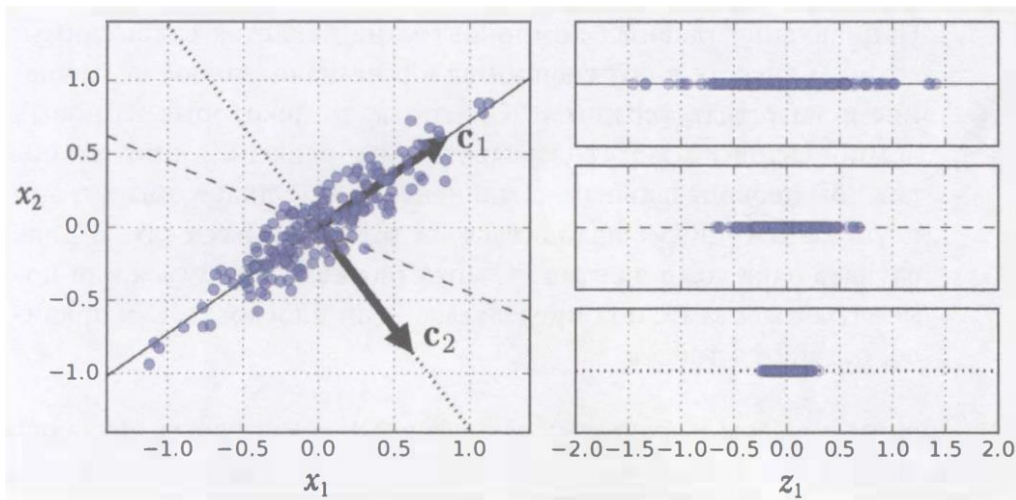

**Fig. 1.** Choosing a hyperplane for projection

The correct choice will be the axis that preserves the maximum dispersion, since it is likely to lose less information compared to other projections. In other words, this axis will minimize the mean square of the distance between the original dataset and its projection onto the axis. This is the idea behind the PCA algorithm.

The algorithm identifies the axis along which the greatest variance is produced in the training set. This axis, or rather the unit vector that represents it, will be referred to as the first component. Then the algorithm finds the second axis, perpendicular to the first, which produces the largest residual dispersion. The third component will be a vector perpendicular to both axes found in the previous steps, and preserves the maximum value of the residual variance. In Figure 1, vector $c_1$ is the first component, and $c_2$ is the second.

After identifying all the main components, we can then reduce the dataset dimension to d by projecting it onto the hyperplane defined by the first d main components. Using this hyperplane ensures that the projection will preserve the maximum possible dispersion as well as the information content of the transformed data.

### 1.5. Processing text and categorical attributes

Most machine learning algorithms prefer working with numerical data, thus several techniques have been created for converting text data to numerical input.

**Label encoding**

One of these techniques is to encode categorical attributes by matching each category with a unique integer.

In general, this is the only work the encoder has to do. Depending on the data, however, this conversion can create a new problem. A connection exists between a set of probably non-related categories after they are converted into a set of numbers due to the possibility of comparing them with each other. Machine learning algorithms will assume that two numerical values close to each other are more similar than two values which are remote from each other.

**One-Hot Encoding**

A common solution to the problem identified above is to create one binary attribute per category: one attribute is set to 1 when the category is the string "hello" (0 otherwise), another at-tribute is 1 when the category is the word "world" (0 otherwise), etc. This technique is called one-hot encoding, as only one attribute value is equal to 1 (hot), while the rest have the value of 0 (cold).

**TF-IDF Vectorizer**

When the task involves classification of text documents, the TF-IDF (Term Frequency - Inverse Document Frequency) algorithm is often used to determine the importance of a word for a specific document relative to other documents. [16]

If a term is used frequently in a certain text, but rarely in others, then it has higher significance for this text. If the term is used quite frequently in all documents in the set, then the significance of such a term will be low after processing.

As the name of the algorithm suggests, the statistical measure used to assess the significance of a word consists of two parts:

1) TF (term frequency), the ratio of the number of occurrences of a word to the total number of words in the document. In this way, the significance of the word $t_i$ within a single document is evaluated.

$$tf(t,d) = \frac{n_t}{\sum_k n_k},$$

where $n_t$ is the number of occurrences of the word $t$ in the document, and the denominator is the total number of words in this document.

2) IDF (inverse document frequency) is the inverse value of the frequency with which a specific word occurs in documents within the collection. Accounting for IDF reduces the weight of commonly used words (in Russian, these would include parts of speech such as interjections, conjunctions, particles, and other noise). There is only one IDF value for each unique word within a particular collection of documents. [17]

$$idf(t, D) = \log \frac{|D|}{|\{d_i \in D | t \in d_i\}|},$$

where |D| is the number of documents in the collection, and $|\{d_i \in D | t \in d_i\}|$ is the number of documents from collection |D|, in which t occurs. [18]

The choice of the logarithm base in the idf formula is insignificant, as changing the base leads to a change in the weight of each word by a constant factor, which does not affect the ratio of weights.

Thus, the TF-IDF measure is the product of two factors:

$$TfIdf(t, d, D) = tf(t, d) \times idf(t, D)$$

## 2. Construction of classifiers

In order to build a machine learning model capable of classifying data with a high level of accuracy, it is necessary to generate a large field of features and select informative data from this dataset.

### 2.1. Feature generation and selection

The methods of feature selection have been studied rather thoroughly, and there are many algorithms for this task. However, feature generation is a rather meticulous process (Fig. 2), more of an art than a science, which involves extracting a set of meaningful features from a large, non-normalized collection of data, most of which tends to carry an excessive amount of information, which should reflect the data structure well and make a positive contribution to the classification of this data.

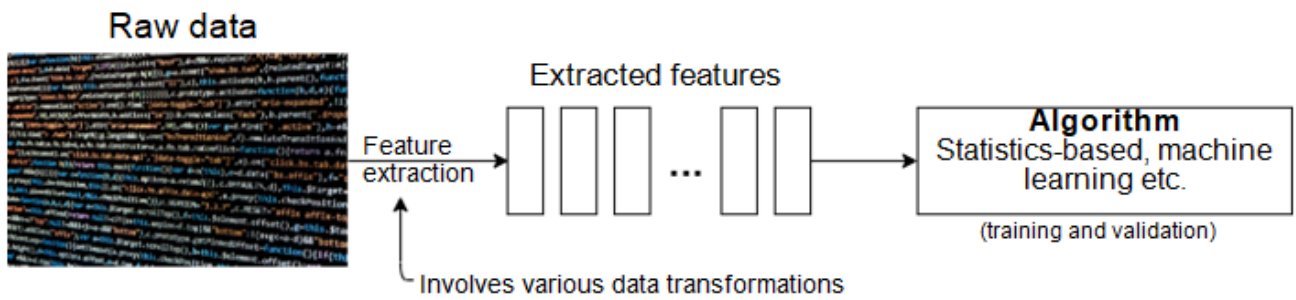

**Fig. 2.** Feature generation process

It is worth noting that this particular task is often the most difficult, and obtaining machine learning algorithms of a very high quality depends on the successful selection and creation of features.

**Categorical features**

Let us assume that objects have a feature that can have values from a specific finite set. For example, a colour may be red, green, blue, or its value may be unknown. In this case, it may be useful to add features such as "unknown", "is_red", "is_green", "is_blue", "is_red_or_blue", and other possible combinations.

**Numeric variables**

If the variable value is a real number, it often helps to round or split it into the entier and fractional parts. You can also convert a numerical feature into a categorical one. For example, if one feature is speed, then other features may be introduced such as the formulations "speed is greater than X", "speed is between X and Y", "speed is greater than Y", which will generalize the feature and increase its information content.

**Aggregated Attributes**

It is practical to add features that aggregate a set of features for a certain object, thereby reducing the dimension of the feature space. As a rule, this is useful in tasks where objects contain several parameters of the same type. For example, a carmaker that produces vehicles with different engine powers. In this case, you can consider features that correspond to high, low, and average engine power to determine which type of cars the company manufactures (e.g. sports or city cars).

**Results obtained using other algorithms**

It is often possible to introduce a result obtained using other algorithms as a feature. For example, if we are dealing with a classification problem, we can first solve the auxiliary clustering problem, and then use the cluster to which the object belongs as a feature in the initial problem. This is usually performed based on the initial data analysis in cases when the objects are well clustered.

**Adding new features**

This point relates to practical tasks from real life. In order to build a high-quality machine learning model, it is often necessary to have subject-specific expertise and an understanding of what affects the target variable. Returning to our example of determining the category of a car, everyone knows that sports cars have far greater engine power than city cars. In a more complex subject area, however, it may be more difficult to arrive at these kinds of conclusions.

### 2.2. Collecting data from an executable file

The first task here is to implement modules for parsing a binary file and extracting data from its structure.

The code is written in Python and uses the "pefile" library to extract data from the structure of an executable file. [19]

To conduct the study, 40,000 legitimate executable files were collected from the Windows operating system, and 40,000 malicious executables were obtained using the company's traffic filtering systems.

Sets of different types of attributes are collected from each file, which are shown in Table 1 with descriptions.

**Table 1.** Attributes collected from the executables

| Attributes | Description |
|---|---|
| dense_features | Numeric attributes such as section sizes, file footprint, number of libraries, number of characters imported, etc. |
| strings | All ASCII strings in the file |
| urls | URLs found in the strings using regular expressions |
| manifest | An .xml document (manifest) containing information about the file name, company, software description, library versions, and other metadata |
| imports | A set of features extracted from the import table in the following format: [library name, the group of functions within this library, function name, presence of suspicion] |
| exports | A set of features extracted from the export table in the following format: [the name of exported function] |
| resources | A set of features extracted from the resource directory table in the following format: [type, name, language] |
| blacklisted strings | A set of strings considered suspicious by malware analysts in the following format: [type, language] |
| tls | This table contains a description of the static variables relating to a thread local storage (TLS). TLS is a Windows-specific method of storing data in which a data object is not a stack variable but is nevertheless local to each separate thread. As a result, each thread has its own instance of data located in TLS. A set of attributes in the following format: |

|  | [AddressOfCallBacks] |
|---|---|
| relocations | Information from the address settings table in the following format: [initial RVA, setting type, block size in bytes] |

### 2.3. The choice of metrics for assessment of classifiers

In the simplest example, the classification quality assessment metric could be the percentage of files which the classifier has classified correctly. This type of metric is referred to as Accuracy in machine learning, and its formula is:

$$Accuracy = \frac{TP + TN}{ALL},$$

where TP is the number of malicious files correctly identified by the classifier, TN is the number of legitimate files correctly identified by the classifier, and ALL is the entire sample. [20]

In order to assess the quality of the classifier, the metrics of precision and recall are introduced separately for each class. The classifier's precision within a class is the percentage of objects that truly belong to a given class in the entire set of objects that the system has assigned to that class. The classifier's recall is the ratio of the number of a given class objects found by the classifier to the entire number of the objects of that class in the test set.

Clearly, the higher the precision and recall of the classifier, the better. In real life however, maximum precision and maximum recall cannot be achieved at the same time, so it is necessary to find an acceptable tradeoff between these two metrics. The F-measure is normally used as a metric that combines information about the precision and recall of an algorithm, the formula for which is:

$$F = (\beta^2 + 1)\frac{Precision \times Recall}{\beta^2 Precision + Recall},$$

where β takes a value in the $0 < \beta < 1$ range if priority is given to precision, and $\beta > 1$ if the priority is given to recall. [21]

As an example, see the graph of the F-measure with the priority given to precision (β=0.5) in Fig. 3.

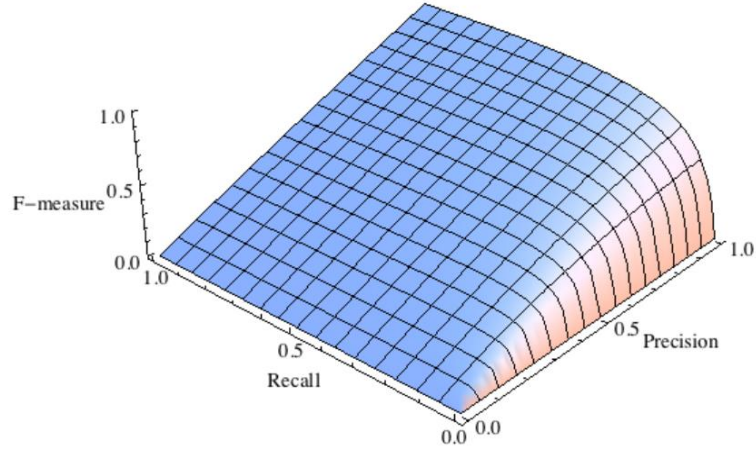

**Fig. 3.** F-measure with the priority given to precision ($\beta^2 = \frac{1}{4}$)

For the specific problem considered in this paper, it is necessary to take into account that the classifier will come across legitimate files in the majority of cases when working with real data. Therefore, it is first and foremost necessary to monitor the recall on the legitimate file class, i.e. to avoid type I errors (false positives). At the same time, the reason the classifier exists is so that malware can be identified. Therefore, the type II error (failure to identify a target) will be used as the second parameter in the final metric. A function follows from the above, which is a combination of type I and type II errors.

The error minimization function, the coefficients in which were selected empirically, is as follows:

$$0.8 \frac{FP}{TN + FP} + 0.2 \frac{FN}{TP + FN} \to min,$$

where
- TN is the number of correctly classified legitimate files,
- FP is the number of false positives (type I errors),
- TP is the number of correctly classified malicious files,
- FN is the number of failures to identify the target (false negatives, type II errors).

### 2.4. Numeric attribute classifier

The first and perhaps the most important classifier is that trained based on numerical features extracted from the binary structure of an executable file and enriched with the features generated when examining the PE file structure and indicators described by malware analysts.

Standard numerical features extracted from an executable file by this module with descriptions are shown in Table 2.

**Table 2.** Standard numerical features of a PE file

| Feature | Description |
|---|---|
| check_sum | File checksum |
| compile_date | Build date |
| debug_size | Debug directory size |
| export_size | Export table size |
| iat_rva | Import table RVA |
| major_version | Major version of the builder |
| minor_version | Minor version of the file |
| number_of_imports | Number of imported libraries |
| number_of_import_symbols | Number of imported symbols |
| number_of_export_symbols | Number of exported symbols (usually subroutines and variables from DLLs) |
| number_of_bound_import_symbols | Number of symbols in the static import binding table |
| number_of_bound_imports | Number of libraries in the import binding table |
| number_of_rva_and_sizes | Number of descriptors and data directories |
| number_of_sections | Number of sections |
| total_size_pe | Physical size of the binary PE file |
| virtual_address | RVA of the first section in memory |
| virtual_size | The size of the first section in memory |
| virtual_size_2 | The size of the second section in memory |
| datadir_IMAGE_DIRECTORY_ENTRY_BASERELOC_size | Physical size of the relocation table |
| datadir_IMAGE_DIRECTORY_ENTRY_RESOURCE_size | Physical size of the resource table |
| datadir_IMAGE_DIRECTORY_ENTRY_IAT_size | Physical size of import address table (IAT) |
| datadir_IMAGE_DIRECTORY_ENTRY_IMPORT_size | Physical size of import table |
| datadir_IMAGE_DIRECTORY_ENTRY_EXPORT_size | Physical size of export table |
| pe_char | File attributes |
| pe_dll | The file is a library (0 or 1) |
| pe_driver | The file is a driver (0 or 1) |
| pe_exe | The file extension is .exe (0 or 1) |
| pe_i386 | The processor is i386 (0 or 1) |
| pe_majorlink | Major version of the builder |
| pe_minorlink | Minor version of the builder |
| size_code | The total size of all sections containing code |
| size_image | The footprint of the file, including all headers |
| size_initdata | The total size of all sections containing initialized data |
| size_uninit | The total size of all sections containing uninitialized data |

We are attempting to train the classifier based on these attributes and test it using the K-fold validation method on five blocks.

Cross-validation is a model validation technique to check how well the model works on an independent dataset using the given statistical analysis and training set. [12] Cross-validation is typically used in forecast problems, and the predictive model would ideally be evaluated in field conditions.

One cross-validation cycle for K blocks involves splitting the dataset into parts, then building the model using one part (called the training set), and validating it on the other part (called the test set). To reduce variation in the results, several cross-validation cycles are performed using different ways of splitting the initial dataset, and the validation results are averaged over all cycles.

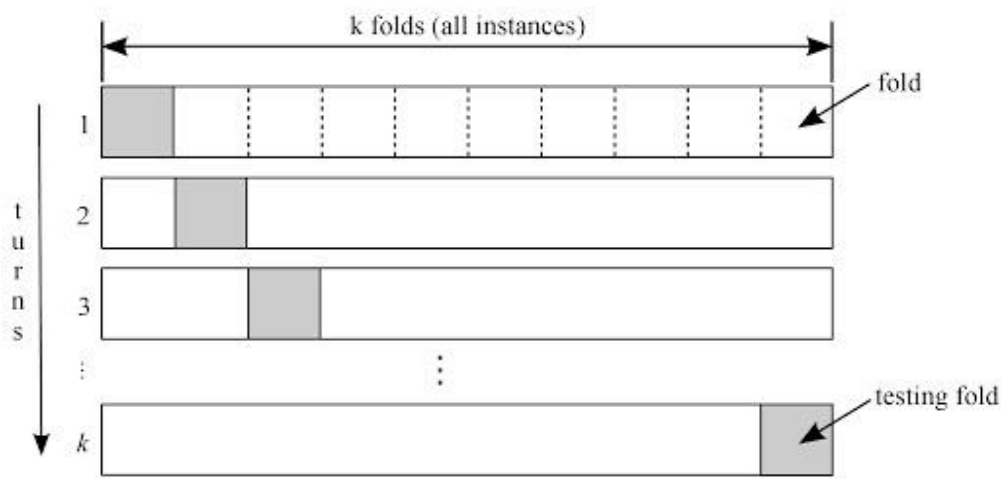

**Fig. 4.** Cross-validation on K blocks

In Figure 4, the initial dataset is divided into K (in this case, K=5) blocks of equal size. One of the K blocks is set aside to test the model, and the remaining K-1 blocks are used as a training set. The process is repeated K times, with each block only used as a test set once. One K result is obtained for each block, which are then averaged or combined in another way to produce a single estimate.

Before training the classifier, we will try to visually evaluate the sample by using the t-SNE dimensionality reduction algorithm [23], which basically models high-dimensional objects with two- or three-dimensional points in a Euclidean space, so that similar objects are modeled with points located close to each other, while dissimilar objects are likely to be modeled with points set far apart from each other. As can be seen from Figure 5, malicious samples mostly overlap with legitimate executables, which indicates an inadequate selection of features.

Let us use the Random Forest algorithm for classification, proven to be the most effective in the field of machine learning. [24] The implementation of this algorithm is available in the scikit-learn machine learning library [25].

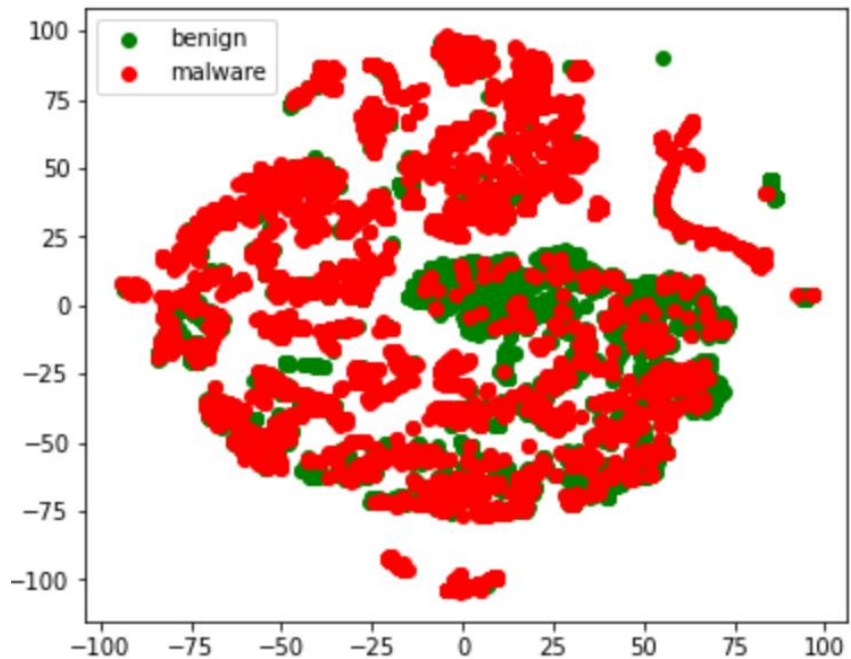

**Fig. 5.** Visualization of standard numerical features of a PE file

An average assessment of the model quality using 5-fold cross-validation according to the metric selected in Section 2.3 yielded 0.08048. In this case, the threshold at which the loss function reaches this minimum is 0.74 (Fig. 6). This means that all samples with forecasts above this threshold value are malicious, and all files with forecasts below this threshold are legitimate.

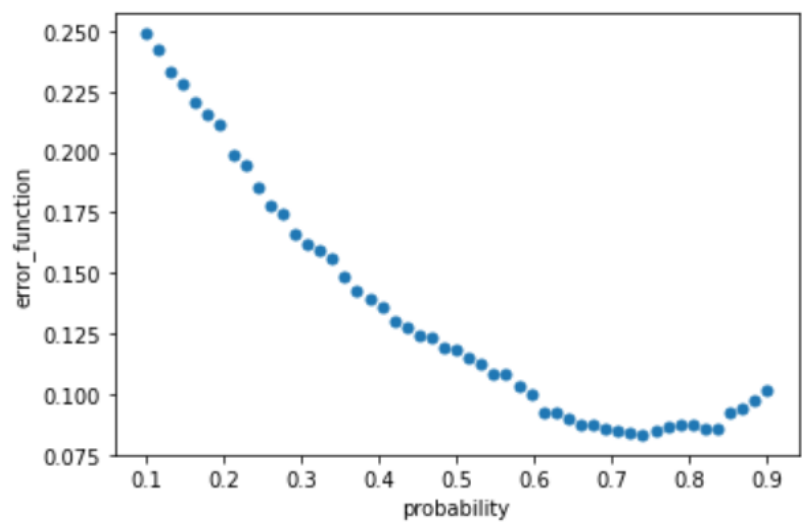

**Fig. 6.** Loss function graph from section 2.3

The selected features show that this dataset has proved to be effective. The recall on malware class files is 0.746 and the recall on benign class files is 0.959 (i.e. 95.9% of all clean files are identified correctly).

The researchers sought to add features that could increase data separability and contribute to the classification. Some of the added features/ groups of features are given in Table 3.

**Table 3.** Generated numeric features of a PE file

| Feature | Description |
|---|---|
| std_sections_names | Whether all section names are standard (.text, .data, .bss, .reloc, .rsrc etc.) |
| generated_check_sum | Generated checksum |
| entropy_sec_mean | Mean entropy of all sections |
| entropy_sec_std | Standard deviation of the entropy of all sections |
| sec_entropy_* | Entropy of standard sections (a name is substituted instead of *, for example sec_entropy_text) |
| known_sections_number | Number of sections with standard names (.text, .data, .bss, .reloc, .rsrc, etc.) |
| unknown_sections_number | Number of sections with non-standard names |
| known_sections_pr | The percentage of sections with standard names out of the total number of sections |
| unknown_sections_pr | The percentage of sections with non-standard names out of the total number of sections |
| *_entropy_sec_mean | Mean entropy for a certain type of section (e.g. writable sections |

| | MEM_WRITE or sections containing executable code CNT_CODE) |
|---|---|
| *_entropy_sec_std | Standard deviation of the entropy for a specific type of section |
| rawsize_sec_mean | Mean size of all sections in a file |
| vasize_sec_mean | Average swap size for all sections |
| std_sections_names | Whether all section names are standard (.text, .data, .bss, .reloc, .rsrc etc.) |
| generated_check_sum | Generated checksum |
| entropy_sec_mean | Mean entropy of all sections |
| entropy_sec_std | Standard deviation of the entropy of all sections |
| sec_entropy_* | Entropy of standard sections (a name is substituted instead of *, for example sec_entropy_text) |
| known_sections_number | Number of sections with standard names (.text, .data, .bss, .reloc, .rsrc, etc.) |
| unknown_sections_number | Number of sections with non-standard names |
| known_sections_pr | The percentage of sections with standard names out of the total number of sections |
| unknown_sections_pr | The percentage of sections with non-standard names out of the total number of sections |
| *_entropy_sec_mean | Mean entropy for a certain type of section (e.g. writable sections MEM_WRITE or sections containing executable code CNT_CODE) |
| *_entropy_sec_std | Standard deviation of the entropy for a specific type of section |
| rawsize_sec_mean | Mean size of all sections in a file |
| vasize_sec_mean | Average swap size for all sections |
| std_sections_names | Whether all section names are standard (.text, .data, .bss, .reloc, .rsrc etc.) |
| generated_check_sum | Generated checksum |
| entropy_sec_mean | Mean entropy of all sections |
| entropy_sec_std | Standard deviation of the entropy of all sections |
| sec_entropy_* | Entropy of standard sections (a name is substituted instead of *, for example sec_entropy_text) |
| known_sections_number | Number of sections with standard names (.text, .data, .bss, .reloc, .rsrc, etc.) |
| unknown_sections_number | Number of sections with non-standard names |
| known_sections_pr | The percentage of sections with standard names out of the total number of sections |

The sample may now be re-visualized using the t-SNE algorithm after reducing the dimensionality of the feature space to two-dimensional Euclidean space. Figure 7 shows that the clusters now stand out much more clearly than in the previous experiment, which indicates that there will be good results at the testing stage for the new classifier.

After training and testing the classifier, the loss function value dropped to 0.06153. The percentage of malware that was correctly detected (malware recall) increased from 74.63% to 88.12%, while the optimal threshold (where the loss function is minimal) moved closer to 0, dropping to 0.628. This indicates a better separation of the sample compared to the previous result.

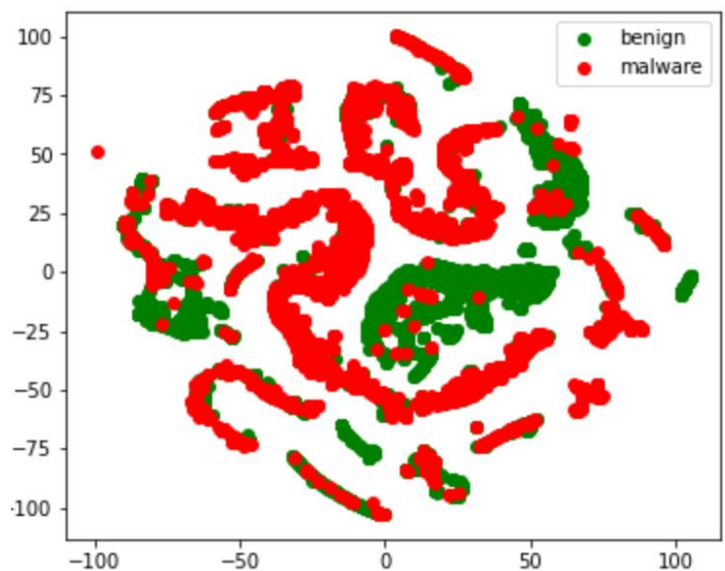

**Fig. 7.** Data visualization including new numerical features

The 50 most informative features selected by the classifier and sorted by importance can be plotted as a histogram. Apart from standard features, such as the date the file was compiled (compile_date) and the size of the data directory with debugging information (debug_size), the newly generated features also make the top 50. The list of leading features includes the generated checksum, the mean entropy of sections with initialized data, the standard deviation of section entropy, the mean footprint of sections with initialized data, and many others.

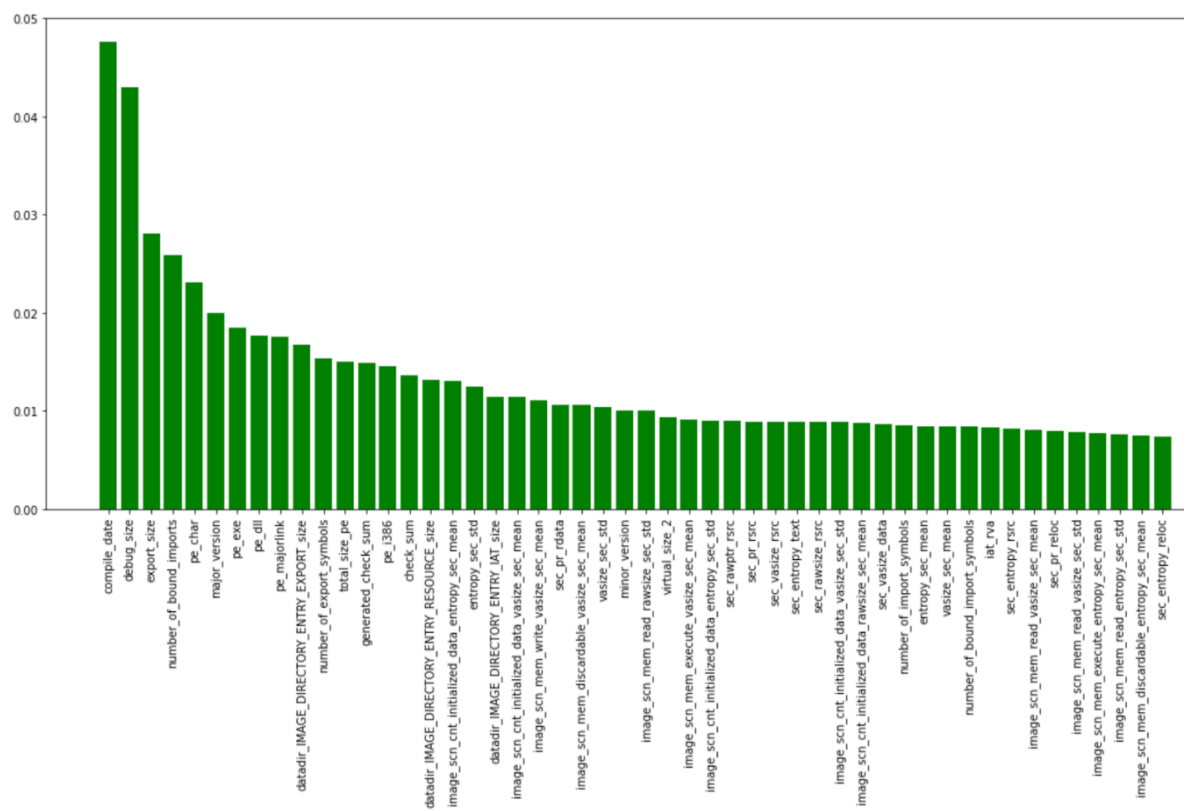

**Fig. 8.** Histogram of the informativeness of numerical features

The trained classifier showed the best result for the selected loss metric (0.05904) on the first 50 significant features shown in Figure 8. The recall on clean files also increased significantly (to 0.961), while we did not sacrifice the share of detecting malicious files from the total number (86.26%).

Looking at the visualization of the new feature space (Fig. 9), it is easy to see that the number of samples creating noise and polluting clusters of other classes has decreased. This allows the conclusion to be made that using strictly informative features eliminates the excess noise that interferes with classification, which enables the classifier to build better forecasts.

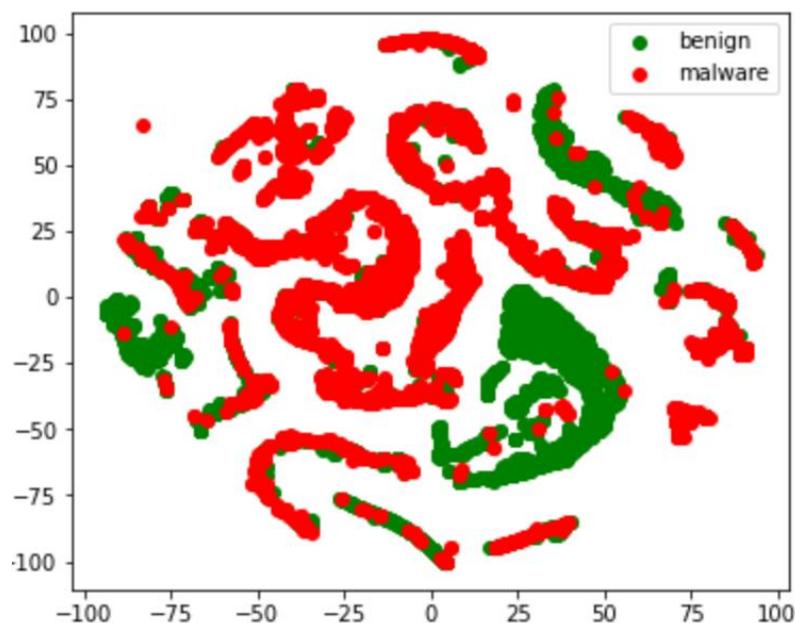

**Fig. 9.** Visualization of the feature space for informative features

Let us add indicators written by malware analysts to the attribute space of our classifier:

- Digital certificate is used which is not valid
- Entrypoint in section <.section_name> which is not executable
- Entrypoint is in last section
- The age of the debug file is suspicious
- The count of imports is suspicious
- The debug file name extension is suspicious

- The dos-stub is missing
- The executable contains some default passwords
- The executable has a lot of executable sections
- The executable has section(s) that are both executable and writable
- The file has been compiled with Delphi
- The file has blacklisted section name(s)
- The file has no Manifest
- The file ignores Address Space Layout Randomization (ASLR)
- The file ignores Data Execution Prevention (DEP)
- The file ignores Structured Exception Handling (SEH)
- The file ignores cookies on the stack (GS)
- The file implements Control Flow Guard (CFG)
- The file imports suspicious number of anti-debug function(s)
- The file is a Device Driver
- The file is code-less
- The file is resource-less
- The file opts for Address Space Layout Randomization (ASLR)
- The file opts for Data Execution Prevention (DEP)
- The file opts for cookies on the stack (GS)
- The file references a debug symbols file
- The file references keyboard functions
- The file references keyboard keys like a Keylogger
- The file-ratio of the resources is suspicious
- The first section is writable
- The last section is executable
- The shared section(s) reached the max threshold
- The size of code is bigger than the size of code sections
- The size of initialized data reached the max threshold
- The size of the optional-header is suspicious

- The value of 'SizeOfImage' is suspicious

After the training, in addition to numerical features, the following indicators (listed in descending order of their informativeness) were included in the top 50 most informative indicators:

1. The file ignores Address Space Layout Randomization (ASLR),
2. The file references a debug symbols file,
3. The file opts for Data Execution Prevention (DEP),
4. The file opts for Address Space Layout Randomization (ASLR),
5. The debug file name extension is suspicious,
6. The file ignores Data Execution Prevention (DEP).

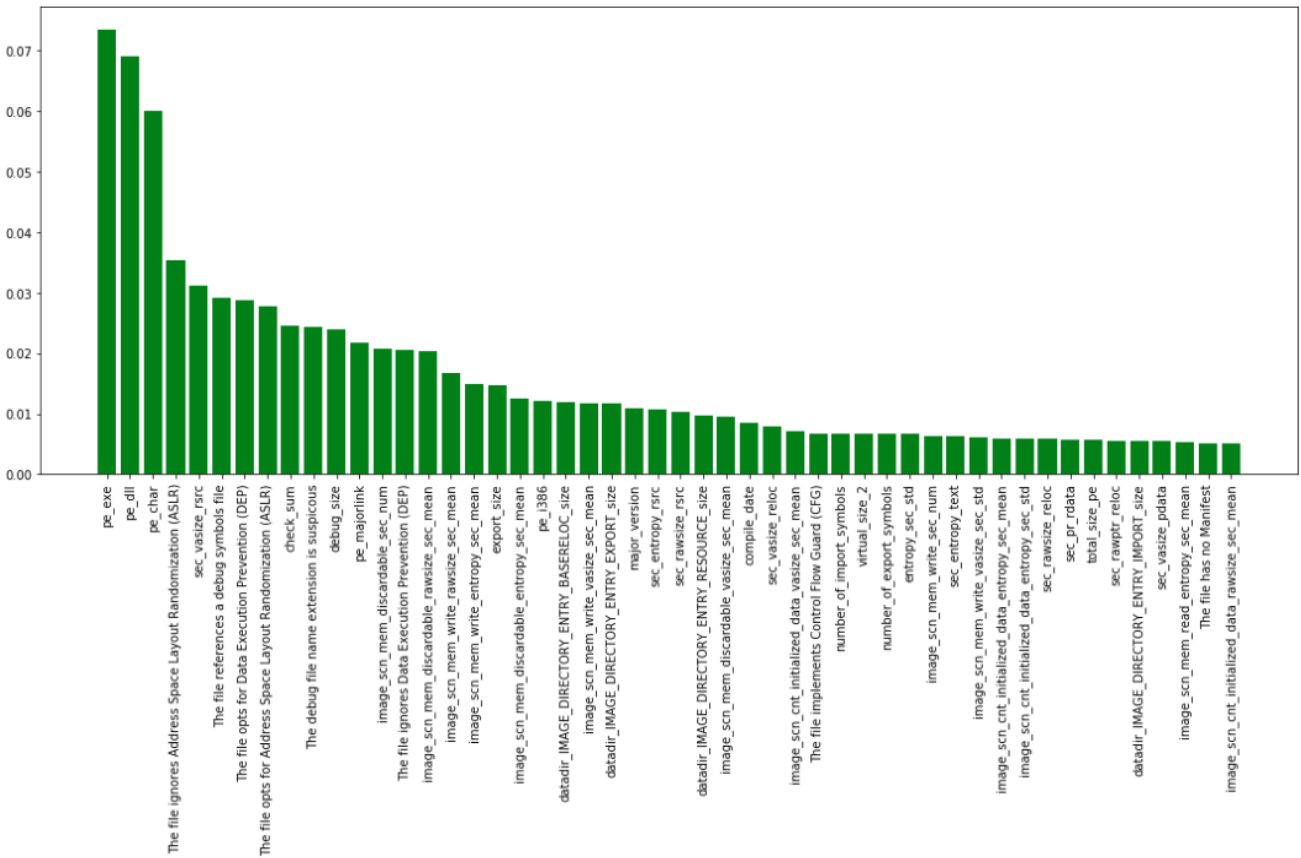

**Fig. 10.** Histogram of the informativeness of numerical features with adding indicators

The following results were obtained during validation of the classifier:
- Recall on clean files: 0.997
- The percentage of correctly identified malicious files increased to 98.53%
- The threshold value is 0.77

- The value of the loss function reached at the selected threshold decreased by more than 25 times compared to the previous results and amounted to 0.002318

The result shows that often, in order to build a high-quality machine learning model, you need to be an expert in a specific field and understand what affects the target metric. In the case being analyzed, the quality of classification was greatly influenced by the indicators described by malware analysts.

### 2.5. String-based classifier

Developing a classifier based on ASCII strings extracted from a binary file involves several stages:

1) ASCII strings are extracted from different areas of the binary file and collected in one array
2) All strings from the array are concentrated into a single string (adding a space be-tween them)
3) A set of such strings is generated for the entire training set
4) The strings are processed using the tf-idf algorithm described earlier using the stop_words dictionary for the English language, and the 5,000 most informative strings are selected. This is done in order to remove noise and speed up model train-ing, as there is often 100,000 or more strings in the binary files.
5) The set of features with the assigned weights (calculated using the tf-idf algorithm) is passed from the scikit-learn library to RandomForestClassifier.

The loss function showed 0.02361 at a threshold of 0.68, which is slightly higher than that of the classifier using numerical attributes. The percentage of malware detection was 94.02%, the completeness on benign files was 0.985.

Let us construct a histogram of the first 50 informative lines (Fig. 11).

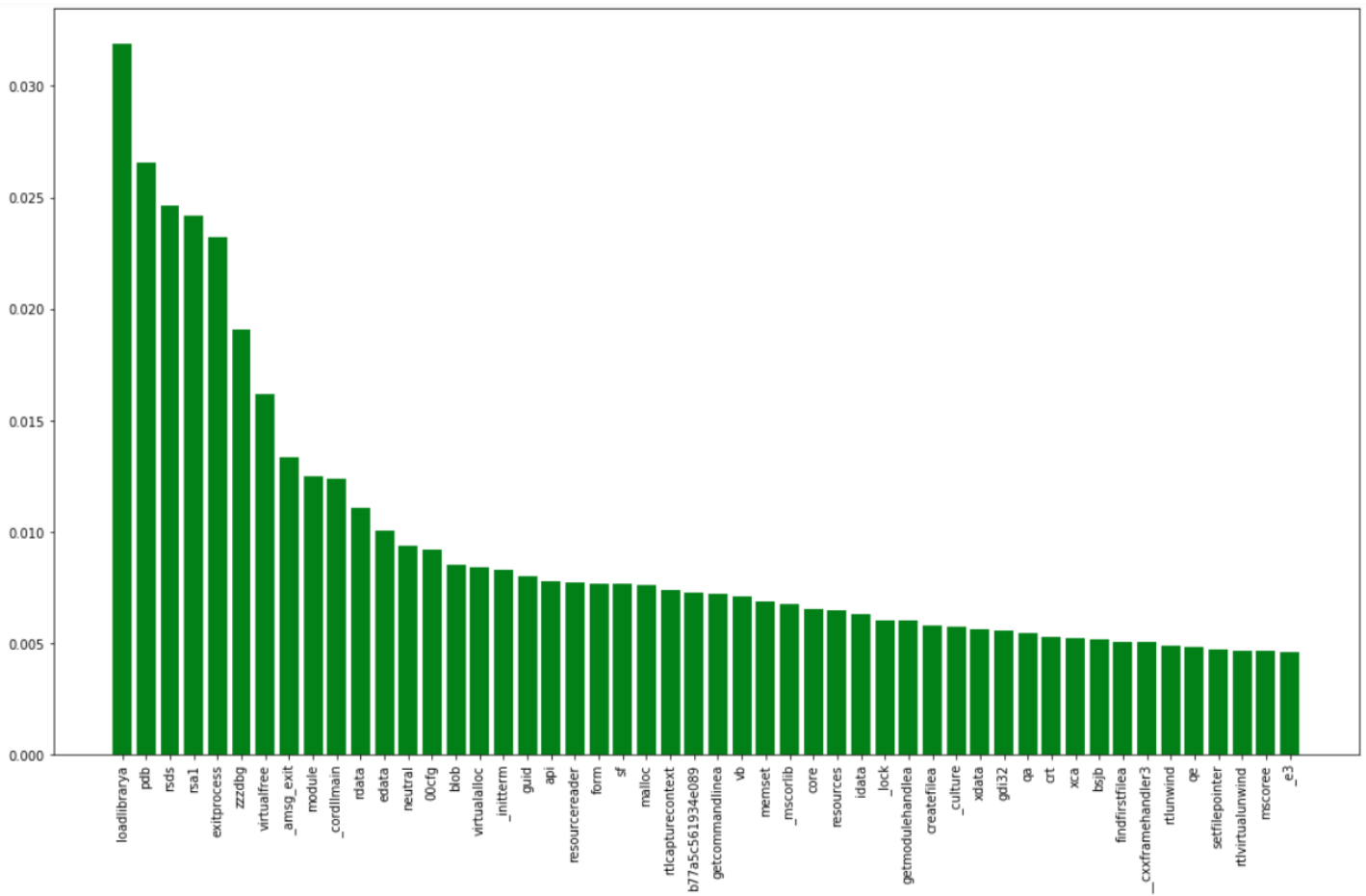

**Fig. 11.** The histogram of string informativeness

The "pdb" string turned out to be the most informative one. This type of file contains the program database, which is a proprietary file format for storing debugging information about a program module. The presence of symbol files (.pdb) on the client machine can be useful, as more detailed information about the exception can be obtained from a .pdb file if an unhandled exception occurs. These files are rarely found in malware, as malware aims to give the virus analyst as little useful information as possible.

Let us test this hypothesis. In order to do so, each of the informative strings will be assigned to one of the classes (Table 4). A string is considered characteristic for a specific class if the ratio of the number of class samples containing this string to the number of all samples containing this string exceeds 0.7. Thus, the "pdb" string identified by the classifier as the most important feature does indeed indicate that the code is legitimate, which confirms that the classifier works.

**Table 4.** Informative strings for each software class

| Malware | Legitimate software |
| --- | --- |

exitprocess, loadlibrarya, getmodulehandlea, gdi32, mg, findfirstfilea, virtualalloc, qa, virtualfree, createfilea, form, aw, kz, ao, lv, getcommandlinea, aj, rtlunwind, vb, qe, fn, vi, ke, sf, createprocessa, jl, jq, setfilepointer, lp, ia, dz, eb, yn, kg, messageboxa, getstartupinfoa, getforegroundwindow, gn, ma, p7, deletefilea, svwuj, fs, wsprintfa, suvw, ar, hn, wininet, kk, jb, og, fw, wc, ec, yp, jg, sn, nz, nm, dispatchmessagea, ow, getcpinfo, lstrcpya, regsetvalueexa, getfileattributesa, getwindowtexta, uj, mw, wa, gettemppatha, 7o, kj, i_, setwindowpos, yo, cb, yx, yg, defwindowproca, fv, qc, nx, qz, bg, ov, nt, gq, zb, jo, xg, comctl32, wq, ki, ox, zw, nq, i7, lb, cz, o2, 6p, qg, postquitmessage, nn, ea, sendmessagea, pppppppp, setstdhandle, getsystemdirectorya, createwindowexa, qk, mt, ga, nc, mp, interlockeddecrement, lstrlena, iswindowenabled, qb, er, oe, ns, ze, ne, 700wp, lstrcmpia, onpf, managed_vector_copy_constructor_iterator, wn, ek, wo, vector_deleting_destructor, za, nw, __ptr64, 6z, bj, uqpxy, settimer, sy, cg, wk, fgiu, f3, zp, sssss, d2, sd, beginpaint, yq, ng, __pascal, wb, interlockedincrement, findnextfilea, placement_delete, getdc, dynamic_initializer_for_, opc, lz, getlastactivepopup.

pdb, zzzdbg, rsa1, _initterm, _amsg_exit, _cordllmain, rsds, module, guid, crt, rtllookupfunctionentry, blob, memset, lsystem, xca, idata, edata, bsjb, malloc, _culture, core, _unlock, neutral, gctl, api, 00cfg, rdata, rtlvirtualunwind, bss, _mscorlib, _purecall, rtlcapturecontext, _except_handler4_common, resources, b77a5c561934e089, xdata, _a_a, 30319, xia, gfids, win, _lock, brc, disablethreadlibrarycalls, __cxxframehandler3, _vsnwprintf, memcpy, 0e3, microsoft_corporation, resourcereader, a_a, mscoree, uvwatauavawh, xiz, uvwh, terminate, watauavawh, _microsoft_corporation, runtimeresourceset, xe3, _e3, 0a_a, dllgetclassobject, lca, pdata, _initterm_e, 8e3, pa_a, yaxxz, xcz, fd9, __c_specific_handler, dllcanunloadnow, _xcptfilter, processthreads, pe3, assemblyproductattribute, assemblyfileversionattribute, _watauavawh, debuggableattribute, padpadp, memmove, _uvwatauavawh, _wcsicmp, _cxxthrowexception, runtime, debuggingmodes, ilist, libraryloader, xcu, _copyright_, ha_a, _h3e_h3e, assemblycopyrightattribute, svwatauavawh, uatauavawh, usvwatauavawh, xiaa.

## 2.6. URL-based classifier

Similarly to the previous classifier, we will build a classifier that can detect malware based on information about URLs stored among other ASCII strings collected from a binary file.

The only difference between the construction of this classifier from the previous one is that this time the tf-idf algorithm analyzes 5-grams of characters extracted from URLs instead of words. Each sample is a string of URLs separated by a space. Note that when an 5-gram overlaps the end of a URL, the 5-gram is padded with white spaces instead of including a part of the next URL.

The 5-fold cross-validation found that the mean minimum value of the loss function achieved at the threshold value of 0.71 is equal to 0.0615. The recall for legitimate files is 0.986, and the recall on malware is 0.901.

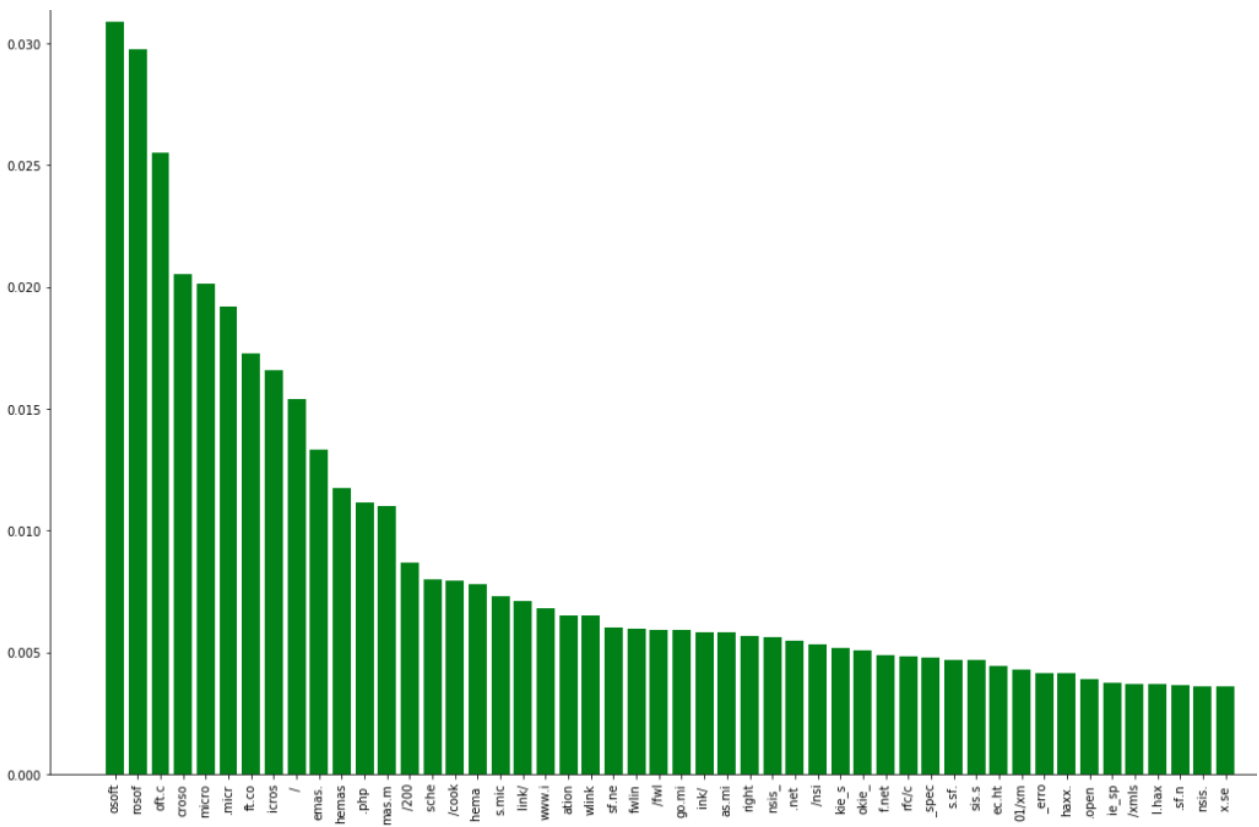

**Fig. 12.** Histogram of the first 50 informative 5-gram URLs

Table 5 lists the n-grams characteristic for a particular type of software.

**Table 5.** Informative n-grams for each software class

| Malware | Legitimate software |
|---|---|
| ` / `, `fc/co`, `.php `, `f.net`, `okie_`, `s.sf.`, `ec.ht`, ` nsis`, ` /rfc`, ` curl`, `.net`, `haxx.`, `s_err`, `x.se `, ` /nsi`, `sis.s`, `is_er`, `.haxx`, `ookie`, `_erro`, `e_spe`, `sis_e`, `/cook`, `rror `, `c.htm`, `_spec`, `curl.`, `ie_sp`, `is.sf`, `url.h`, `c/coo`, `l.hax`, `bsens`, `spec.`, `sf.ne`, `rfc/c`, `nsis_`, `nsis.`, `www.i`, `pec.h`, ` %s `, `.sf.n`, `rl.ha`, `/nsis`, `error`, `ftwar`, `oftwa`, `xx.se`, `tware`, `gnlog`, `ware.`, `axx.s`, `ww.go`, `ww.ib`, `erts/`, `re.co`, `ogle.`, `le.co`, ` /%s `, `kie_s`, `.goog`, `e.php`, `.ibse`, `w.ibs`, `oogle`, `/inde`, ` /pki`, `.crl0`, `ndex.`, `i/crl`, `gle.c`, `ensof`, `te.ph`, `index`, `lp/in`, `softw`, `gin.r`, ` crl.`, `a.com`, `in.ru`, `%s:80`, ` %s:8`, `pki/c`, `are.o`, `jrsof`, `googl`, `ownth`, `c.com`, `.crt0`, `%s:%d`, `/ishe`, `/rfc/`, `ate.p`, `are.c`, `s:80 `, `n.ru `, `dmin/`, `raut_`, `arch.`, `that.`, `senso`, `elp/i`, `w.goo`, `cerau`, `ishel`, ` butt`, `rl/pr`, `at.co`, ` %s:%`, `cooki`, `gate.`, `.exe `, `nthat`, `ll/ga`, `s/mic`, `-06-2`, `ibsen`, ` i2.t`, `rsoft`, ` gnlo`, `p/ind`, `/gate`, `butte`, `ki/ce`, `www.j`, `hoco.`, `ex.ph`, `.com0`, `min/b`, ` /ish`, `.jrso`, `crt0 `, `shelp`, `etmas`, `dex.p`, `n/bul`, `wntha`, | `oft.c`, `icros`, `.micr`, `croso`, `ft.co`, `osoft`, `rosof`, `micro`, `emas.`, `go.mi`, ` /fwl`, ` go.m`, `fwlin`, `hema`, `wlink`, `mas.m`, `/fwli`, ` /200`, `ation`, `s.mic`, `link/`, `ink/ `, `as.mi`, `o.mic`, `hemas`, `hub.c`, `.com)`, `sche`, `/xml/`, `ithub`, `githu`, `/winf`, `pport`, `names`, ` ' `, `thub.`, `xamar`, `nfx/2`, `l/199`, `/xmls`, ` /dot`, `/1998`, `2000/`, `amesp`, `otnet`, `/core`, ` loca`, `fx/20`, `xmlsc`, `mlsch`, `mespa`, `marin`, `/xaml`, `ml/19`, `host `, `w.xam`, `/2001`, `x/200`, `ocalh`, `com) `, `xml/1`, `senta`, `ub.co`, `calho`, ` /xml`, `01/xm`, `trans`, ` gith`, `in.co`, `.xama`, `ww.xa`, `uppor`, `dotne`, `lhost`, `/2000`, `espac`, `006/x`, `8/nam`, `refx/`, `p.org`, `2001/`, `amari`, `98/na`, `xaml `, `alhos`, `a.ms`, `prese`, `1/xml`, `tree/`, `/pres`, `.avir`, `resen`, `local`, `rin.c`, `vira.`, `lsche`, `aka.m`, `nses/`, `001/x`, `orefx`, `suppo`, `/dotn`, ` aka.`, `06/xa`, `licen`, `t/cor`, `cense`, `es/gp`, `tnet/`, `999/x`, `ty.co`, `/name`, `avira`, `www.x`, `/2006`, `ntati`, `ity.c`, `00/xm`, `l.htm`, `enses`, |

`re.or`, `lc.co`, `i/cer`, `ts/mi`, `nsoft`, `2011_`, `ww.jr`, `x.php`, `_2010`, `ocsp.`, `.down`, `253.6`, `0-06-`, `o.net`, `tm0@`, `ercho`, `w.php`, `%d%s`, `utter`, `ww.ne`, `w.net`, `bi.d`, `choco`, `3.crl`, `rchoc`, `tterc`, `2010-`, `53.6`, `ersll`, `d.php`, `nlogi`, `w.jrs`, `3.crt`, `llc.c`, `124.2`, `odesi`, `.netm`, `crl0z`, `/rpa`, `124.`, `s:%d%`, `ocsp`, `aut_2`, `h.php`, `rl.mi`, `4.217`, `ingca`, `ank/d`, `i.dow`, `.253.`, `dupe.`, `:%d%s`, `.com:`, `7.253`, `bull/`, `/cps0`, `ia.co`, `crl0a`, `tmast`, `11_20`, `in/bu`, `t.php`, `sters`, `/bull`, `.site`, `cure.`, `_2011`, `l/pro`, `/domo`, `terch`, `opca2`, `27:33`, `eraut`, `011-1`, `ull/g`, `etea.`, `gca.c`, `upe.p`, `co.ne`, `mping`, `eyout`, `sign.`, `out.p`, `rl0z`, `/lsb`, `rsllc`, `ki/cr`, `om:44`, `/lsba`, `czban`, `6-23.`, `com0`, `sllc.`, `/cps`, `load.`, `tampi`, `u.com`, `ets.c`, `/admi`, `bi.do`, `p.ver`, `24.21`, `crl.v`, `ayme.`, `i3.t`, `bank.`, `how.p`, `w.ver`, `ww.ve`, `/mcy.`, `/dupe`, `pc/ch`, `x.htm`, `gn.co`, `eb2a.`, `rpa0`, `ts.co`, `gamet`, `money`, `33355`, `netma`, `.baid`, `ign.c`, `/nbb`, `d.com`, `tss-c`, `oco.n` `mlns/`, `servi`, `arin.`, `compa`, `net/c`, `ka.ms`, `et/co`, `ernam`, `6/xam`, `.acro`, `name`, `s.org`, `lns/`, `x/tre`, `s/gpl`, `icens`, `space`, `ction`, `/win`, `i.org`, `port`, `/tree`, `000/0`, `m.org`, `pace`, `right`, `xmlns`, `u.org`, `ffice`, `l/pre`, `1998/`, `rname`, `-comp`, `ressi`, `/exp`, `ml/pr`, `mlfor`, `nu.or`, `coref`, `gpl.h`, `.dtd`, `winfx`, `openx`, `link`, `ses/g`, `ibili`, `esent`, `logi`, `xaml/`, `a.net`, `/lice`, `direc`, `9/xsl`, `nuget`, `efx/t`, `/lic`, `ports`, `ority`, `in.wi`, `gnu.`, `nd/20`, `irect`, `gnu.o`, `forma`, `et.or`, `get.o`, `sl/tr`, `/xmln`, `fx/tr`, `ww.gn`, `ssues`, `lsoap`, `ira.c`, `.gnu.`, `ldsig`, `as.op`, `commo`, `09/xm`, `bug.c`, `ratio`, `00/09`, `ows.n`, `lp/>`, `/net`, `eserv`, `eskto`, `uget.`, `port.`, `pl.ht`, `/xsl/`, `.nuge`, `infor`, `tatio`, `nslat`, `a.org`, `ww.av`, `ws.ne`, `sions`, `sktop`, `offic`, `nfigu`, `essio`, `ww.co`, `eview`, `patib`, `xmlso`, `nform`, `/sql/`, `ww.ad`, `/09/x`, `form`, `/sql`, `/con`, `ware/`, `ectio`, `998/n`, `oap.o`, `illa.`, `ra.co`, `lity/`, `d/200`, `xmlfo`, `.wind`, `issue`

**2.7. Classifiers based on categorical attributes**

Another type of classifier is a classifier based on categorical attributes. In cases where each attribute is a feature of a particular category, the most practical approach is to encode each category's set of attributes using the one-hot encoding technique discussed earlier. This will produce a vector of features, each of which is equal to 1 if this attribute is present in the file, and 0 if that is not the case.

This section will describe the construction of such a classifier based on data from the resource directory table. This classifier displayed the best results out of all the rest (classifiers based on manifests, on the import and export tables, on data from tls and relocations, and on strings that are considered suspicious were also built and tested).

**Classifier based on data from the resource directory table**

The classifier is developed as follows:

1. The feature sets are extracted from the resource table: names of libraries, library function groups (if the group is unknown, the value is set to "undefined"), the names of the imported functions and whether a given function is suspicious (information about suspicious functions was taken from commercial pestudio software [26]).

2. Each feature is sorted according to the set of feature types, and is included in the corresponding vector.

3. Vectors with feature sets are encoded using the "one-hot encoding" technique described earlier.

4. All data is run through the pre-trained OneHotEncoder class and converted into vectors, each element of which encodes an attribute. This element contains a number, which represents the number of times this attribute was encountered in the file.

5. Feature space is reduced using the PCA algorithm, preserving 99% of the original variance. This way, excess noise is removed, classifier training speed is increased, and the memory footprint reduced.

6. The training dataset with reduced dimensionality of features is passed to RandomForestClassifier.

Initially, the resulting feature dimensionality for a training dataset of 2010 elements was equal to 15150, and after dimensionality reduction preserving 99% of the initial dispersion, it dropped to just 58.

It should be noted that at this stage, extraction of informative features is a futile process, as the classifier did not work with the original features, but with their compacted presentation. The only thing that can be done is a visual evaluation of the data quality by projecting the attribute space onto a two-dimensional plane using the t-SNE algorithm (Fig. 13).

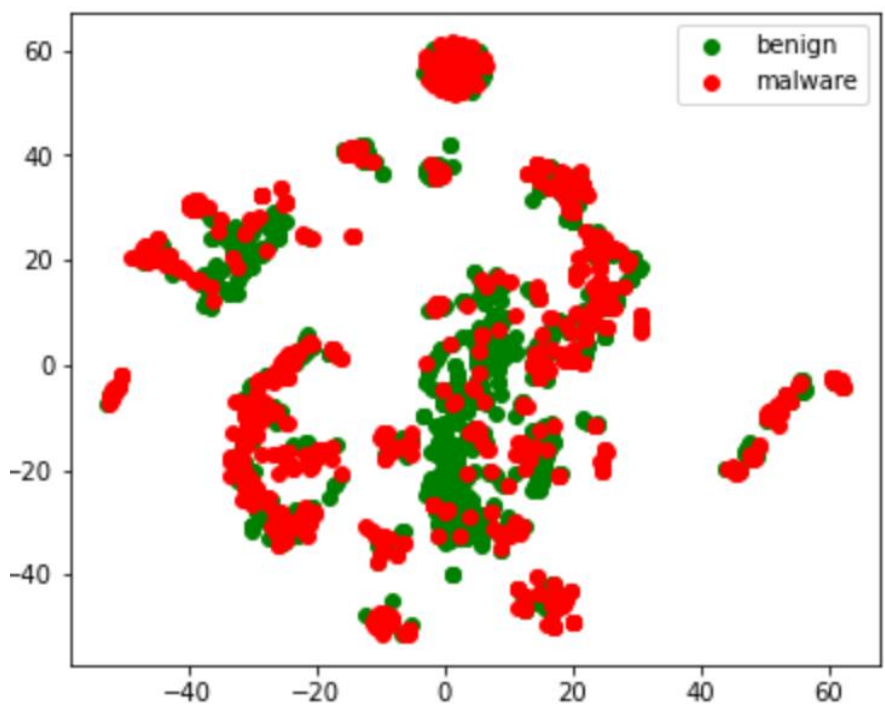

**Fig. 13.** Feature space visualization for the import table categorical attributes

The 5-fold cross-validation showed that the mean minimum value of the loss function is equal to 0.03226, which is much better than that of other models built on the same principle but using different data. This value is achieved with a threshold value of 0.80. The recall on legitimate files is 0.986, and percentage of malware detection is 89.94%.

### 3. Constructing an ensemble

It is widely acknowledged that the winners of many machine learning competitions have used several ensemble methods (the most famous ones are from the competition [27]).

An analogy can be used to explain why ensemble learning is better than single models. Suppose we asked thousands of random people a question, and then aggregated their answers. In many cases, this type of aggregated answer is better than an answer provided by a person with subject-specific expertise. This is called the wisdom of the crowd. Similarly, if we aggregate the forecasts given by a group of several different predictors (which is called an ensemble), we will often get better forecasts than those obtained from the best individual predictor [28]. Moreover, if we aggregate forecasts from predictors, each of which is tailored to its own feature space, then a much larger and more informative feature space will be covered than the feature space of each individual predictor.

Guided by this idea, we can build a classifier that combines all the classifiers that have been built before it. It will aggregate the forecasts of all classifiers and predict the class with the highest class probability averaged over all individual classifiers.

This method is called a soft voting classifier. This ensembling approach often achieves greater efficiency than hard voting, whereby the class that received the largest number of votes is chosen (hard voting classifier), because more weight is given to votes from classifiers with high confidence.

The weight of each classifier in the constructed ensemble is inversely proportional to the value of the loss function achieved at the stage of testing the classifier, and normalized so that the sum of the weights of all classifiers gives 1:

$$\frac{(1 - loss\_function\_value)^{10}}{\sum_{i=0}^{9}(1 - loss\_function\_value_i)^{10}}$$

**Table 6.** Weights of trained classifiers in the ensemble

| dense       | 0.1443 |
|-------------|--------|
| strings     | 0.1355 |
| urls        | 0.0990 |
| resources   | 0.1152 |
| imports     | 0.1233 |
| manifest    | 0.0744 |
| relocations | 0.0965 |
| blacklisted | 0.0772 |
| tls         | 0.1345 |

If there is a classification failure for a particular classifier (e.g. no URL was found in the PE file), the ensemble does not take this classifier into account when making the final decision.

The ensemble was validated using a dataset consisting of 19,100 clean files and 676,000 malicious files. Among the malicious files, 611,434 were correctly identified, the refusal to classify (when for some reason it was not possible to open/extract data from the PE-file) amounted to 55,790 malicious files, and the goal skips were, respectively, 8,776. Among the clean files, 28 files were false positives, which is 0.15 of all clean files in the dataset.

The training resulted in achieving better performance indicators compared to individual classifiers. The recall on legitimate files reached 0.9985, and the recall on malicious files increased to 0.9858, which is higher than that of the strongest classifier in the ensemble using numerical attributes of the executable file.

## Conclusion

This paper described in detail the complete process of creating machine learning models capable of detecting malware with a high level of precision in a non-isolated environment, based on static analysis of a binary PE file. This avoids running the software for dynamic research, which is typical for dynamic malware analysis systems in a sandbox.

A method for representing a binary PE file was chosen. A number of data processing techniques have been proposed for data extracted from the executable file that enable efficient classification. A feature engineering method has been applied, in which a number of features are generated to increase classification precision, based on information about the executable file structure and analysis of malicious code.

An ensemble of classifiers was created on the basis of the well-known ensemble learning approach [29], which uses weighted aggregation of forecasts from individual classifiers, each of which is tailored to its own feature space. As a result, a much larger feature space is covered than each individual classifier could have.

In turn, a machine learning system can be built based on a trained ensemble of classifiers, which can detect malicious code before it is run.

The code for the software developed over the course of the study can be viewed in the Colaboratory interactive development environment by following the link [31].